\documentclass[letter,twocolumn]{jpsj3}
\usepackage{txfonts}
\usepackage{color}

\title{
Numerical Diagonalization Study of the Phase Boundaries 
of the $S=2$ Heisenberg Antiferromagnet on the Orthogonal Dimer Lattice 
}
\catcode`\@=11
\def\simle{\mathrel{\mathpalette\@versim<}}   
\def\simge{\mathrel{\mathpalette\@versim>}}   
\def\@versim#1#2{\lower2.5pt\vbox{\baselineskip0pt \lineskip-.5pt
   \ialign{$\m@th#1\hfil##\hfil$\crcr#2\crcr\sim\crcr}}}
\catcode`\@=12

\author{Hiroki Nakano$^1$, 
T\^oru Sakai$^{1,2}$,  
and Yuko Hosokoshi${}^{3}$}
\inst{$^1$
Graduate School of Science, 
University of Hyogo,
Kamigori, 
Hyogo 678-1297, Japan \\
$^2$
National Institutes for Quantum and Radiological Science and Technology, 
SPring-8
Sayo, Hyogo 679-5148, Japan \\
${}^{3}$Department of Physics, Osaka Metropolitan University, Osaka 
558-8585, Japan 
}

\abst{
The $S=2$ Heisenberg antiferromagnet 
on the orthogonal dimer lattice 
is studied. 
The edges of the exact dimer and N$\acute{\rm e}$el-ordered phases 
in the ground state of the system 
are examined by the numerical diagonalization method. 
Our present results are discussed 
by combining them with previously obtained estimates 
for smaller-$S$ cases. 
We find that an intermediate region 
between the exact dimer and N$\acute{\rm e}$el-ordered phases 
gradually widens as spin $S$ is increased up to $S=2$. 
}

\begin{document}
\maketitle


The Heisenberg antiferromagnet on the orthogonal dimer lattice has 
garnered significant attention as a frustrated magnets, 
similar to the kagome and triangular lattice antiferromagnets. 
This orthogonal dimer system is referred to 
as the Shastry--Sutherland model\cite{Shastry_Sutherland_Physica1981}. 
The frustration often gives rise 
to various phase transitions and exotic quantum states. 
However, notably, 
the number of quantum states that can be obtained
in a mathematically rigorous form is limited. 
In this situation, the Shastry--Sutherland model serves 
as a valuable example, wherein the exact dimer ground state is realized 
when the orthogonal dimer interactions are sufficiently strong.  
Conversely, the model is in the N$\acute{\rm e}$el-ordered phase 
when the interactions that form the square lattice 
are adequately strong. 
The majority of subsequent studies has concentrated on the case of $S=1/2$.
The discovery of SrCu$_2$(BO$_3$)$_2$ 
as a good candidate material in Ref.~\ref{Kageyama_PRL1999} 
has  notably increased the importance of the $S=1/2$ model.  
Extensive and intensive studies have been conducted 
on this system from theoretical and 
experimental perspectives\cite{Albrecht1996,SMiyahara_KUeda_PRL1999,
WeihongPRB1999,AKoga_NKawakami_PRL2000,YFukumoto_JPSJ2000,
Chung_PRB2001,Lauchli_PRB2002,Lou_arXiv1212_1999,
Corboz_Mila_PRB2013,Wang_Batista_PRL2018}.  
These studies have led to a widespread consensus that 
SrCu$_2$(BO$_3$)$_2$ is in the dimer phase 
of the Shastry--Sutherland model. 

However, research on cases in which spin $S$ is greater than 1/2 is scare. 
Shastry and Sutherland\cite{Shastry_Sutherland_Physica1981} demonstrated that 
the direct product of the two spin states 
that form a singlet dimer is a rigorous eigenstate of the system, 
not only in the $S = 1/2$ case but also in larger-$S$ cases.  
When the dimer interaction $J_{1}$ is sufficiently strong, 
this exact dimer eigenstate is the ground state of the system 
with the system size $N$ having the energy 
\begin{equation}
E_{\rm ED}=-\frac{1}{2}NJ_{1}S(S+1) .
\label{exact_dimer_energy}
\end{equation}
In the case of strong interactions forming the square lattice $J_{2}$, 
the N$\acute{\rm e}$el-ordered phase emerges irrespective of $S$.  
According to Kanter\cite{Kanter_PRB_1989},  
a rigorous region of the ratio of the interactions 
as a necessary condition as follows: 
\begin{equation}
\frac{J_{2}}{J_{1}/2} \le \frac{1}{S+1} , 
\label{Kanter_boundary_S}
\end{equation}
for $S \ge 1$ and 
\begin{equation}
\frac{J_{2}}{J_{1}/2} \le 1  , 
\label{Kanter_boundary_spin_one_half}
\end{equation}
for $S = 1/2$ 
when the exact singlet dimer state is the ground state.  
However, this study has been followed by only a limited number 
of numerical investigations\cite{Koga_JPSJ2023,HNakano_JPCM2024}. 
Reference~\ref{Koga_JPSJ2023} focused on the $S=1$ case 
and studied the anisotropy effect in the system; 
only a single-sized sample including 16 spins was treated in this study. 
In Ref.~\ref{HNakano_JPCM2024}, for the cases of $S=1$ and 3/2, 
the edges of the exact dimer and N$\acute{\rm e}$el-ordered phases 
for the isotropic case were determined 
based on calculations for 16- and 20-site clusters.  

Returning to the $S=1/2$ case, it was pointed out that 
as the $J_{2}$ interaction is increased 
beyond the edge of the exact dimer phase, 
a plaquette singlet state appears 
before the N$\acute{\rm e}$el-ordered phase 
occurs\cite{AKoga_NKawakami_PRL2000}.  
Theoretical attempts to reliably estimate phase boundaries 
and deepen our understanding of the system have been undertaken.  
A few years ago, Ref.~\ref{HNakano_SSL_JPSJ2018} showed that 
the internal structure of the plaquette singlet phase is 
complex and not uniform.  
Subsequently, a series of theoretical studies 
concerning the $S=1/2$ case 
were conducted\cite{Lee_PRX2019,TShimokawa2021,Yang_PRB2022,
TSakai_LT29_JPSCP2023,HNakano_LT29_JPSCP2023,Liu_PRB2024}. 
High-pressure experiments successfully changed 
the ratio $J_{2}/J_{1}$ and reached the boundary of the dimer phase. 
\cite{HOhta_JPhysChem2015,Zayed_NatPhys2017,TSakurai_JPSJ87_ESR_HP}. 

Under the context, the aim of the present letter is 
to clarify the edges of the exact dimer 
and N$\acute{\rm e}$el-ordered phases 
in the $S=2$ case of the Shastry--Sutherland model 
using a computational and unbiased approach.  
We thereby deepen our understanding of the systematic behavior 
of the target system from the exact dimer phase 
to the N$\acute{\rm e}$el-ordered phase.  
This is the purpose of the present study. 


\begin{figure}[tb]
\begin{center}
\includegraphics[width=8cm]{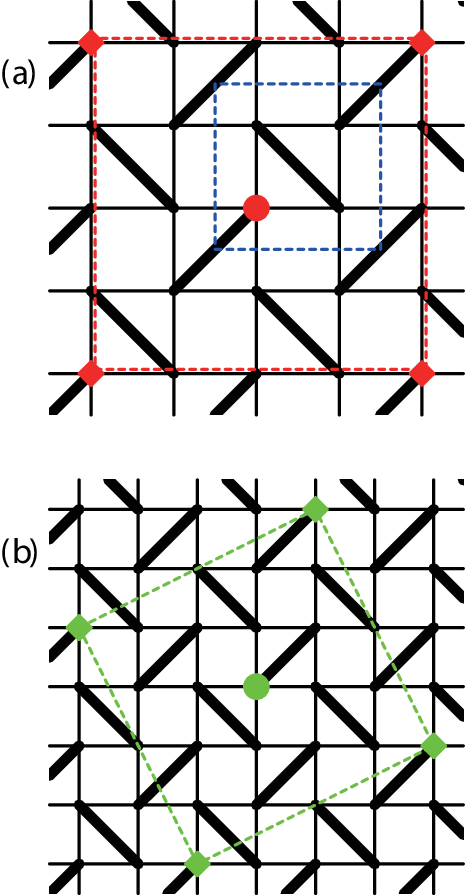}
\end{center}
\caption{(Color) 
Lattice structure of the orthogonal dimer system and 
finite-sized clusters under the periodic boundary condition. 
Thick solid lines show the bonds of the orthogonal dimer interaction ($J_{1}$). 
Thin solid lines represent the interaction forming the square lattice ($J_{2}$). 
The case of $N=16$ is shown by the red broken square in Panel (a) 
and the unit cell ($N=4$) is represented by the blue broken square. 
Note that 
the red broken square is displayed with a slight translational shift 
to avoid situations in which overlapping lines make visibility difficult. 
In Panel (b), the case of $N=20$ is shown using the green broken lines. 
The present study investigates the case when $S=2$ spins are located 
at each vertex of the square lattice, 
whereas in Ref.~\ref{HNakano_JPCM2024}, 
$S=1$ or $S=3/2$ spins are located at each vertex. 
The filled circle in each panel illustrates 
the center site of the square; 
filled diamonds illustrate the corners of the square. 
The pair of the circle and the diamond sites 
is the longest-distant one in each cluster 
under the periodic boundary condition. 
}
\label{fig1}
\end{figure}

The Hamiltonian studied here is given by 
\begin{equation}
{\cal H}
=
\sum_{\langle i ,j\rangle : \ {\rm orthogonal~dimer}} J_{1} 
\mbox{\boldmath $S$}_{i}\cdot\mbox{\boldmath $S$}_{j} 
+
\sum_{\langle i ,j\rangle : \ {\rm square~lattice}} J_{2} 
\mbox{\boldmath $S$}_{i}\cdot\mbox{\boldmath $S$}_{j} 
. 
\label{Hamiltonian}
\end{equation}
Here, $\mbox{\boldmath $S$}_{i}$ 
denotes the $S=2$ spin operator at site $i$. 
In this study, we consider the case of an isotropic interaction 
in the spin space. 
Site $i$ is assumed to represent a vertex of the square lattice. 
The number of spin sites is represented by $N$. 
The first term of Eq.~(\ref{Hamiltonian}) expresses 
the orthogonal dimer interactions illustrated 
by thick solid bonds in Fig.~\ref{fig1}. 
The second term of Eq.~(\ref{Hamiltonian}) expresses 
the interactions forming the square lattice
illustrated by thin solid bonds in Fig.~\ref{fig1}. 
This study considers that the two interactions between the two spins 
are antiferromagnetic, 
namely, $J_{1} > 0$ and $J_{2} > 0$. 
Energies are measured in units of $J_{1}$. 
Hereafter, we set $J_{1}=1$. 
We use $r$ to represent the ratio $J_{2}/J_{1}$. 
When $r=0$, the system is an assembly of isolated dimerized spin models. 
However, in the limit $r\rightarrow\infty$, 
the system is reduced to the $S=2$ Heisenberg antiferromagnet 
on the ordinary square lattice.  
This study examines the ratios of the edges of the exact dimer
and N$\acute{\rm e}$el-ordered phases 
denoted by $r_{c1}$ and $r_{c2}$, respectively.  

We examine finite-sized clusters under the periodic boundary condition. 
This study treats the $N=16$ and 20 finite-sized clusters 
illustrated in Fig.~\ref{fig1}. 
Note that $N/4$ is an integer 
because
the number of spins in 
a unit cell of the present system, 
which is illustrated in Fig.~\ref{fig1}(a), is four. 
Note that these two clusters are regular squares. 
These regular square clusters help us 
to capture the essence of the two dimensions of the present system. 

We carry out numerical diagonalizations 
using the Lanczos algorithm to obtain the lowest energy of ${\cal H}$ 
in the subspace belonging to $\sum _j S_j^z=M$. 
Note that $z$-axis is taken as the quantized axis of each spin.  
The prevailing consensus is that 
numerical diagonalization calculations are unbiased.
Consequently, reliable information about the system can be obtained. 
The energy is represented by $E(N,M)$, 
where $M$ is an integer between $-NS$ and $NS$.  
We mainly focus on the case of $M=0$ 
because the ground state energy $E_{\rm g}$ is given by $E(N,M=0)$.  
Some of our Lanczos diagonalizations were carried out 
using MPI-parallelized code that was originally 
developed in a study of Haldane gaps\cite{HNakano_HaldaneGap_JPSJ2009}. 
The usefulness of our program was confirmed via large-scale 
parallelized calculations\cite{
HN_TSakai_S2HaldaneGap_JPSJ2018,
HNakano_JPSJ2019,HNakano_S1HaldaneGap_JPSJ2022}. 
Note that 
the dimension of the matrix of the largest-scale calculations 
in this study is 5,966,636,799,745 in the subspace $M=0$ for $N=20$. 
The calculations have been carried out on Fugaku using 65,105 nodes. 
To the best of our knowledge, 
Ref.~\ref{HN_TSakai_S2HaldaneGap_JPSJ2018} was the first study 
in which a numerical diagonalization 
for an $S=2$ Heisenberg antiferromagnet of $N=20$ size was carried out 
to investigate the Haldane gap issue for a one-dimensional system. 
However, the computational cost of the $S=2$ system with $N=20$ 
still remains substantial even when using Fugaku.  
Our program was confirmed to succeed in treating systems 
with even larger matrix dimensions 
(12,663,809,507,129 in Ref.~\ref{HNakano_JPSJ2019}, 
18,252,025,766,941 in Ref.~\ref{HNakano_S1HaldaneGap_JPSJ2022}, 
and
32,247,603,683,100 in Ref.~\ref{HNakano_LT29_JPSCP2023}) using Fugaku.  
Because the dimension for the $S=2$ case of $N=24$, which is 
the next larger target of the present study, is more than $10^{15}$,
its numerical diagonalization calculations are difficult 
in the present computational environment. 


\begin{figure}[tb]
\begin{center}
\includegraphics[width=8cm]{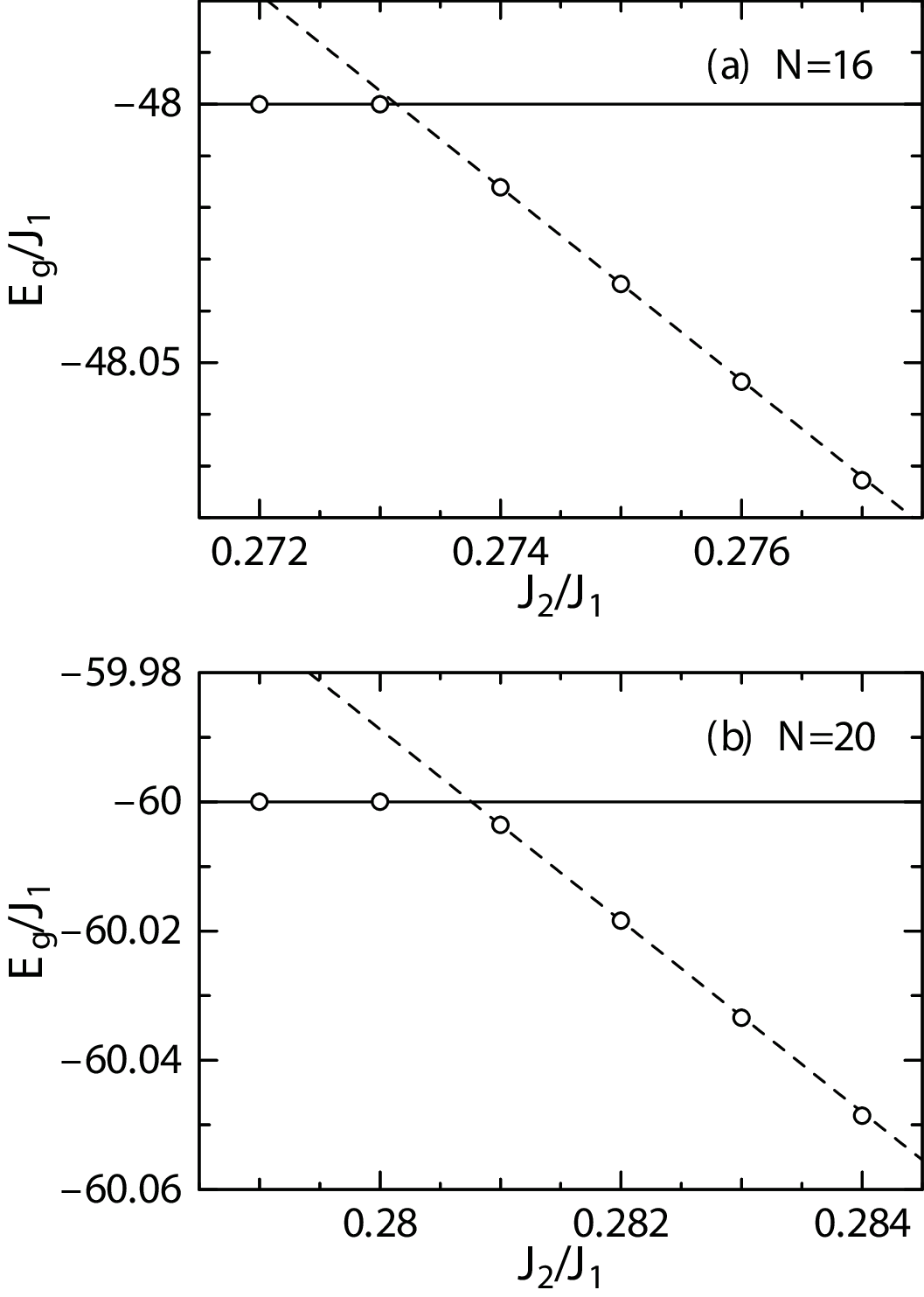}
\end{center}
\caption{
Ground state energy of the $S=2$ orthogonal dimer lattice antiferromagnet 
near the edge of the exact dimer phase. 
Circles represent our numerical diagonalization results 
for $N=16$ and 20 in panels (a) and (b), respectively. 
The horizontal solid line shows the energy of the exact dimer state. 
The broken line in each panel is obtained by fitting 
based on the two points that are close to the energy level 
of the exact dimer state.
}
\label{fig2}
\end{figure}

First, we observe an $r$-dependence of $E_{\rm g}$ 
near the edge of the exact dimer phase 
in the results depicted in Fig.~\ref{fig2}.  
Panels (a) and (b) correspond to the $N=16$ and 20 cases, respectively. 
In each panel, one can observe that 
our calculations successfully capture the energy level ($E_{\rm g}/J_{1}=-3N$)
of the exact dimer state in the region of small $r$. 
In the region where $r$ becomes larger than a specific value, 
$E_{\rm g}$ becomes lower than the exact dimer state. 
In each panel, 
we draw a fitting line determined from two data points 
which are close to the energy level of the exact dimer state. 
Note that other large-$r$ data points that are not used for the fitting 
fall on the fitting line. 
Our calculations strongly suggest that 
a spin state that is different from the exact dimer state exists  
in the range of large $r$ in Fig.~\ref{fig2}. 
The intersection of the horizontal solid line and the broken fitting line 
provides information about the edge of the exact dimer phase.
Our results for the intersection are
\begin{equation}
r = 0.2731 , 
\label{rc1-16s}
\end{equation}
for $N=16$ and 
\begin{equation}
r =
0.2807 , 
\label{rc1-20s}
\end{equation}
for $N=20$. 
The difference between the two results (\ref{rc1-16s}) and (\ref{rc1-20s})
is small. 
These results suggest that the edge of the exact dimer phase is 
\begin{equation}
r_{c1} =0.28(1) .
\label{rc1-final}
\end{equation}
Subsequent analyses will compare this result (\ref{rc1-final}) 
with previously reported results for smaller-$S$. 

\begin{figure}[tb]
\begin{center}
\includegraphics[width=8cm]{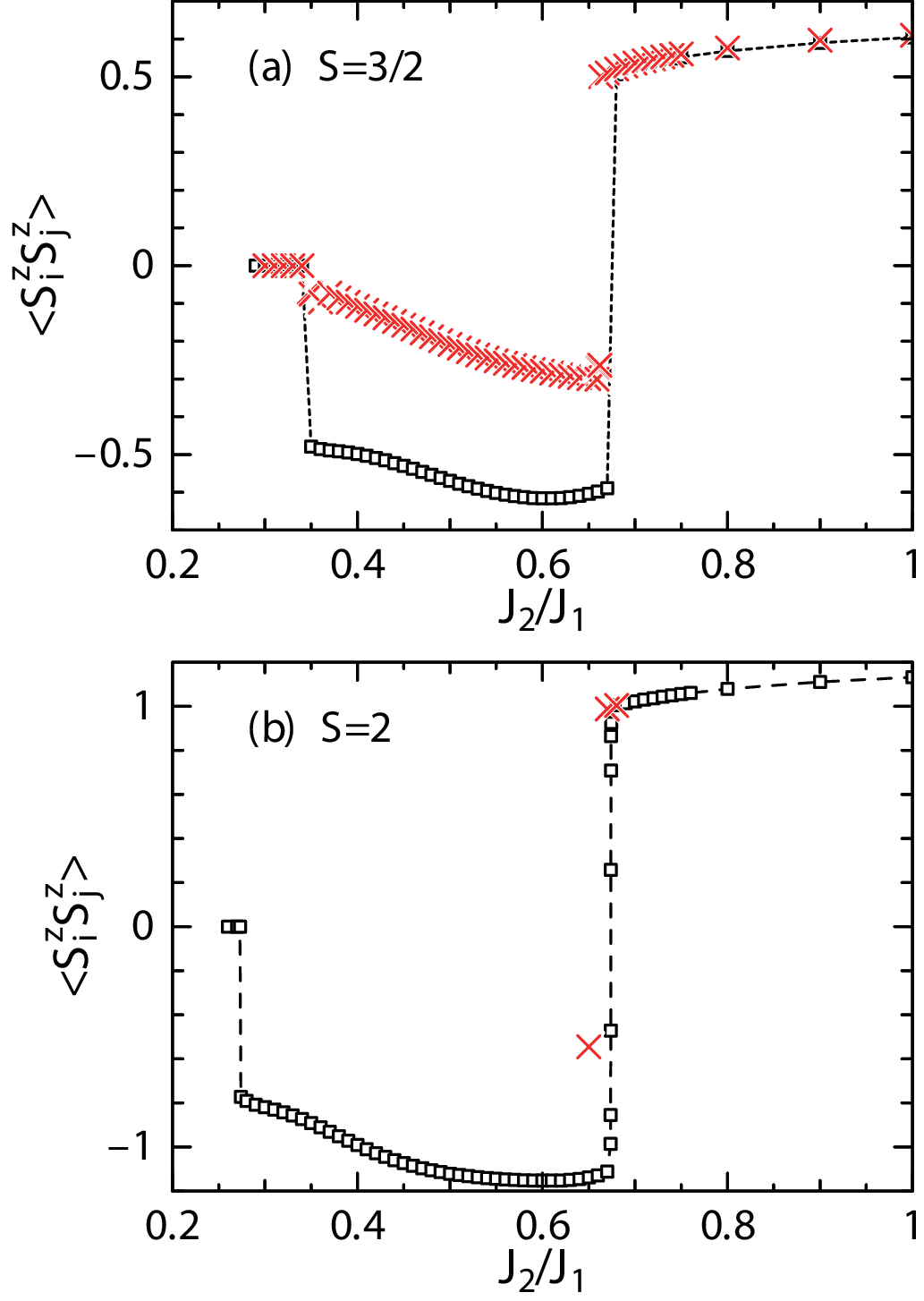}
\end{center}
\caption{(Color) 
Correlation function $\langle S_{i}^{z} S_{j}^{z} \rangle$ 
for the pair of $i$ and $j$ with the longest distance 
in each finite-sized cluster. 
Squares and crosses denote the results for $N=16$ and 20, respectively. 
Panels (a) and (b) show the results for $S=3/2$ and 2, respectively. 
Panel (a) is reproduced from Ref.~\ref{HNakano_JPCM2024} 
so that readers can easily identify the similarities and differences  
between the $S=3/2$ and $S=2$ cases. 
}
\label{fig3}
\end{figure}

Next, we examine the edge of the N$\acute{\rm e}$el-ordered phase 
by observing the spin correlation functions in the ground state. 
We focus on 
$\langle S_{i}^{z} S_{j}^{z} \rangle$ 
for a pair of $i$ and $j$ of the longest distance in the finite-sized clusters. 
This quantity helps us to know 
whether the system exhibits the N$\acute{\rm e}$el order. 
The longest distance is then determined 
as the distance from the center of the regular square 
to each corner of the regular square, 
as illustrated in Fig.~\ref{fig1}. 
Analysis of the correlation function for the longest-distance pair 
was also carried out in a study based 
on Monte Carlo simulations\cite{Sandvik_PRB2010}. 
The present results are depicted in Fig.~\ref{fig3} 
together with the results of $S=3/2$ in Ref.~\ref{HNakano_JPCM2024} 
for comparison. 
The results for $N=16$ over the entire range are presented.  
Above $r\sim 0.68$, 
$\langle S_{i}^{z} S_{j}^{z} \rangle$ takes values that are significantly 
larger than unity. 
When $r$ is decreased from $r\sim 0.68$ to $r\sim 0.67$, 
$\langle S_{i}^{z} S_{j}^{z} \rangle$ rapidly but continuously decreases. 
It also shows discontinuous behavior at the ratio expressed 
in Eq.~(\ref{rc1-16s}). 
However, for $N=20$, 
we concentrate on carrying out our large-scale calculations 
only for the cases for $r=0.65$, 0.67, and 0.68 
because the corresponding computational jobs consume 
many node-hour products as computational resources. 
One finds that 
$\langle S_{i}^{z} S_{j}^{z} \rangle$ of $N=20$ for $r=0.67$ and 0.68 
take values that agree with 
$\langle S_{i}^{z} S_{j}^{z} \rangle$ for $N=16$ above $r \sim 0.68$. 
On the other hand, 
$\langle S_{i}^{z} S_{j}^{z} \rangle$ for $N=20$ for $r=0.65$ takes 
a negative value; 
its absolute value is approximately 50\% 
of the absolute value of $\langle S_{i}^{z} S_{j}^{z} \rangle$ of $N=16$. 
This behavior is shared with the case of $S=3/2$ shown in Fig.~\ref{fig3}(a). 
Therefore, our calculations strongly suggest that 
the edge of the N$\acute{\rm e}$el-ordered phase is located at 
\begin{equation}
r_{c2} =0.66(2) .
\label{corr-final}
\end{equation}
The result will next be compared with previously reported results 
for smaller-$S$.  

\begin{figure}[tb]
\begin{center}
\includegraphics[width=8cm]{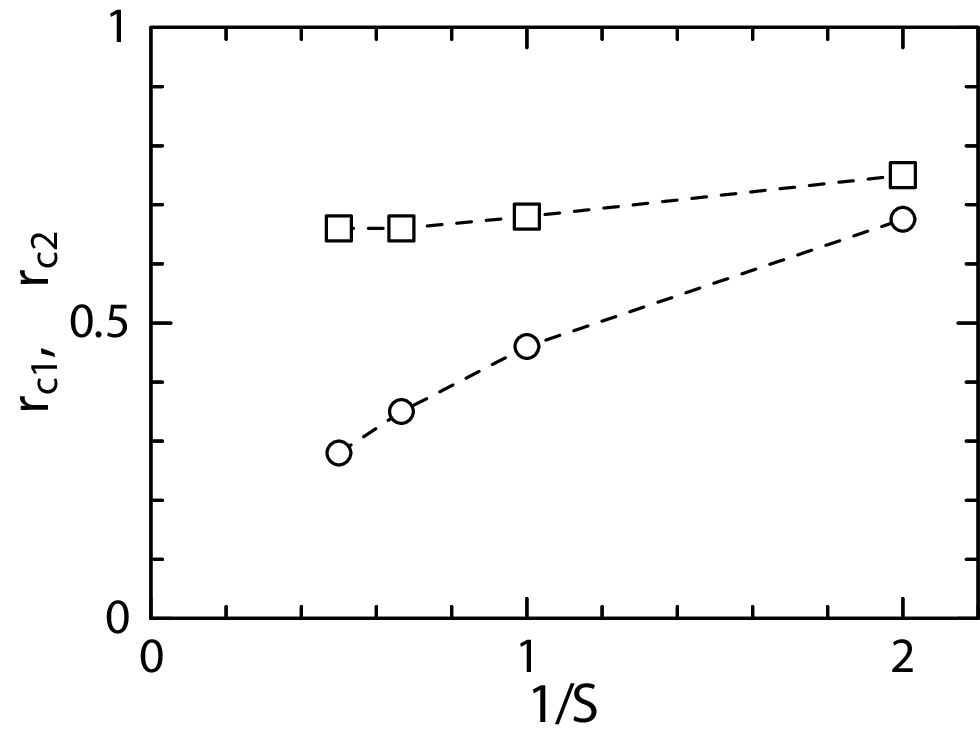}
\end{center}
\caption{
$S$-dependence of $r_{\rm c1}$ and $r_{\rm c2}$. 
Circles and squares denote the results for $r_{\rm c1}$ and $r_{\rm c2}$, 
respectively, 
where $r_{\rm c1}$ and $r_{\rm c2}$  are the edges 
of the exact dimer and N$\acute{\rm e}$el-ordered phases. 
}
\label{fig4}
\end{figure}
Now, let us combine the present results for $S=2$ 
and the results for smaller-$S$ cases reported previously 
to capture the systematic behaviors of $r_{\rm c1}$ and $r_{\rm c2}$ 
as functions of $1/S$. 
The results are depicted in Fig.~\ref{fig4},  
where the data for $S=1$ and 3/2 are taken from Ref.~\ref{HNakano_JPCM2024}, 
the data of $r_{\rm c1}$ for $S=1/2$ from Ref.~\ref{HNakano_LT29_JPSCP2023}, 
and the data of $r_{\rm c2}$ for $S=1/2$ from Ref.~\ref{HNakano_SSL_JPSJ2018}. 
One can observe that $r_{\rm c1}$ gradually decreases as $S$ is increased 
and that $r_{\rm c1}$ seems to vanish 
within the limit of $S\rightarrow\infty$. 
Our present result for the region of the exact dimer ground state 
is significantly wider than the region expressed
using Inequality~(\ref{Kanter_boundary_S}). 
On the other hand, the decrease for $r_{c2}$ with increasing $S$ seems 
monotonic and extremely small.  
Even in the limit of $S\rightarrow\infty$, 
$r_{c2}$ appears to remain nonzero. 
The boundaries of the exact dimer and N$\acute{\rm e}$el-ordered phases 
were examined in Ref.~\ref{Chung_PRB2001}, where 
the author treated a generalized Hamiltonian to Sp(2$n$) symmetry 
and argued for properties under the large-$n$ limit. 
Reference~\ref{Chung_PRB2001} reported that 
the boundary of the ($\pi,\pi$) long-range order phase 
is almost fixed at $J_{2}/J_{1}=1$. 
This fixed boundary is similar to our results showing
that a small decrease in $r_{c2}$ occurs with increasing $S$.  
However, our present result for N$\acute{\rm e}$el-ordered phase 
is considerably wider than that in Ref.~\ref{Chung_PRB2001}. 
In addition, our present result for the region of the exact dimer ground state 
is also wider than that of the dimer short-range order phase 
in Ref.~\ref{Chung_PRB2001}. 
Therefore, the present results for the intermediate region 
between the exact dimer and N$\acute{\rm e}$el-ordered phases 
is narrower than that of the ($\pi,q$) long-range order phase 
between the ($\pi,\pi$) long-range order and the dimer short-range order phases
in Ref.~\ref{Chung_PRB2001}. 
Our narrower intermediate region based on an unbiased 
Lanczos-diagonalization study of the SU(2) Hamiltonian 
suggests that 
the ground state energy of the spin-$S$ Heisenberg antiferromagnet 
on the orthogonal dimer lattice 
in the intermediate region 
is not as low as the corresponding energy 
of the generalized Hamiltonian in the large-$n$ limit.

\begin{figure}[tb]
\begin{center}
\includegraphics[width=8cm]{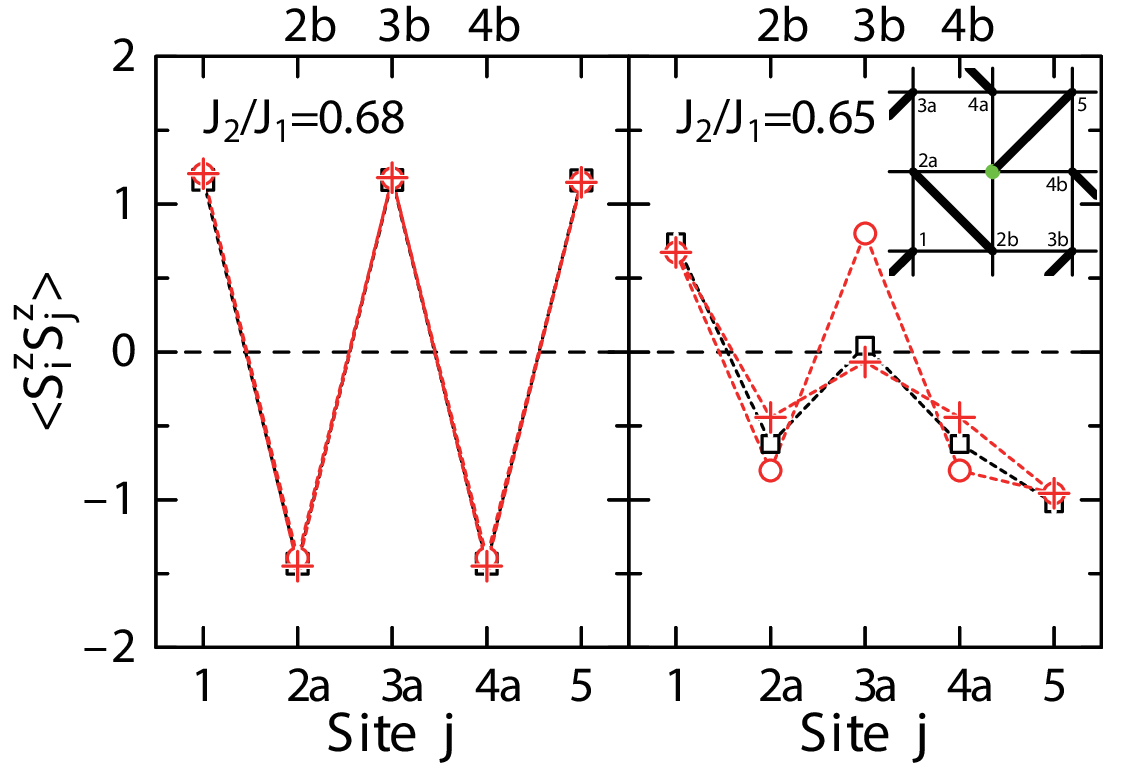}
\end{center}
\caption{
(Color)
Correlation function $\langle S_{i}^{z} S_{j}^{z} \rangle$ 
for the pair of an arbitrary site $i$ and $j$ that is within 
short-range distance. 
The inset shows the numbering of site $j$ 
around site $i$ illustrated by a filled green circle. 
The results for $N=16$ are plotted as black open squares 
and those for $N=20$ are plotted as red open circles and pluses, 
for $j=2{\rm a}, 3{\rm a}, 4{\rm a}$ and $j=2{\rm b}, 3{\rm b}, 4{\rm b}$, 
respectively. 
}
\label{fig5}
\end{figure}
Finally, to capture the properties of the spin state 
of the system in the intermediate region,  
let us observe $\langle S_{i}^{z} S_{j}^{z} \rangle$ 
for the pair of $i$ and $j$ that are close to each other. 
The results are depicted in Fig.~\ref{fig5}. 
Note that the numerical results for the finite-sized cluster with $N=20$ 
show difference between
$j=1\rightarrow 2{\rm a}\rightarrow 3{\rm a}\rightarrow 4{\rm a}\rightarrow 5$
and 
$j=1\rightarrow 2{\rm b}\rightarrow 3{\rm b}\rightarrow 4{\rm b}\rightarrow 5$
because of the tilt of the square of $N=20$ 
although no difference occurs in the $N=16$ case without tilting. 
For $J_2/J_1=0.68$, one can clearly observe a staggered behavior 
in our numerical data, namely, 
$\langle S_{i}^{z} S_{j}^{z} \rangle$ are negative 
for $j=2{\rm a}, 2{\rm b}, 4{\rm a}$, and 4b, 
whereas
$\langle S_{i}^{z} S_{j}^{z} \rangle$ are all positive 
for $j=1, 3{\rm a}, 3{\rm b}$, and 5.  
The staggered orientitaions of spins are related to the result 
of $\langle S_{i}^{z} S_{j}^{z} \rangle$ for the longest-distant pair 
for $J_2/J_1=0.68$ as shown in Fig.~\ref{fig3}(b). 
By contrast, all the results of $|\langle S_{i}^{z} S_{j}^{z} \rangle|$ 
for $J_2/J_1=0.65$ become smaller than those for $J_2/J_1=0.68$. 
For $J_2/J_1=0.65$, the results for $\langle S_{i}^{z} S_{j}^{z} \rangle$ 
for $j=2{\rm a}, 2{\rm b}, 4{\rm a}$, and 4b 
as a nearest-neighbor pair in the square lattice show negative values. 
For $j=5$, $\langle S_{i}^{z} S_{j}^{z} \rangle$ becomes negative, 
which suggests that the $J_2$ interactions disturb 
the staggered orientation of spins. 
Between $j=3{\rm a}$ and 3b, a significant difference emerges for $N=20$. 
For $j=1$, the positive values of $\langle S_{i}^{z} S_{j}^{z} \rangle$ are 
maintained from $J_2/J_1=0.68$ to  $J_2/J_1=0.65$. 
Consequently, the staggered nature survives for $J_2/J_1=0.65$ 
only within the range up to the nearest-neighbor pair. 
Our investigation does not thoroughly clarify properties 
of the ground state in the intermediate region. 
Characteristics of the system in the intermediate region 
should be examined carefully in future. 


In summary, 
we have studied 
the $S=2$ Heisenberg antiferromagnet on the orthogonal dimer lattice. 
The edges of the exact dimer phase $r_{\rm c1}$ and 
N$\acute{\rm e}$el-ordered phase $r_{\rm c2}$ 
have been obtained by applying the Lanczos diagonalization method  
to finite-sized clusters with system sizes $N=16$ and 20. 
Our results for $r_{\rm c1}$ and $r_{\rm c2}$ 
suggest that
an intermediate region appears between the two phases, 
which is wider than those for smaller-$S$ cases. 
Properties of the system in the intermediate regin 
will be further investigated in future studies. 
Such studies will greatly advance 
our fundamental understanding of frustrated magnetism. 

\begin{acknowledgment}

We wish to thank Professor N.~Todoroki for
fruitful discussions. 
This research was partly supported by KAKENHI 
(Grants Nos. 20K03866, 23K11125, 23K25824, and 25K07229). 
In this research, we used the computational resources of the
supercomputer Fugaku provided by RIKEN through the HPCI System Research
projects (Project IDs: hp220043, hp230114, hp230532, hp230537, and hp250164). 
Some of the computations were performed 
using the facilities of the Institute for Solid State Physics, 
The University of Tokyo and 
Supercomputing Division, Information Technology Center, 
The University of Tokyo. 
This research partly used computational resources of Pegasus 
provided by Multidisciplinary Cooperative Research Program 
in Center for Computational Sciences, University of Tsukuba.

\end{acknowledgment}

%
%
%

\end{document}